\documentclass[%
 aip,
 amsmath,amssymb,
 reprint,%
]{revtex4-1}

\usepackage{graphicx}
\usepackage{dcolumn}
\usepackage{bm}
\usepackage{xcolor}

\usepackage[utf8]{inputenc}
\usepackage[T1]{fontenc}
\usepackage{mathptmx}
\usepackage{etoolbox}
\usepackage{comment}
\makeatletter
\def\@email#1#2{%
 \endgroup
 \patchcmd{\titleblock@produce}
  {\frontmatter@RRAPformat}
  {\frontmatter@RRAPformat{\produce@RRAP{*#1\href{mailto:#2}{#2}}}\frontmatter@RRAPformat}
  {}{}
}%
\makeatother
\begin{document}

\preprint{AIP/123-QED}

\title[]{Compressible turbulent convection: The role of temperature-dependent thermal conductivity and dynamic viscosity}
\author{John Panickacheril John}
\affiliation{Institut f\"ur Thermo- und Fluiddynamik, Technische Universit\"at Ilmenau, D-98684 Ilmenau, Germany}
\email{john.panickacheril-john@tu-ilmenau.de}

\author{J\"org Schumacher}%
\affiliation{Institut f\"ur Thermo- und Fluiddynamik, Technische Universit\"at Ilmenau, D-98684 Ilmenau, Germany}
\affiliation{Tandon School of Engineering, New York University, New York City, NY 11201, USA} 
\email{joerg.schumacher@tu-ilmenau.de}

\date{\today}

\begin{abstract}
The impact of variable material properties, such as temperature-dependent thermal conductivity and dynamical viscosity, on the dynamics of a fully compressible turbulent convection flow beyond the anelastic limit are studied in the present work by two series of three-dimensional direct numerical simulations in a layer of aspect ratio 4 with periodic boundary conditions in both horizontal directions. One simulation series is for a weakly stratified adiabatic background, one for a strongly stratified one. The Rayleigh number is $10^5$ and the Prandtl number is 0.7 throughout this study. The temperature dependence of material parameters is imposed as a power law with an exponent $\beta$. It generates a superadiabaticity $\varepsilon(z)$ that varies across the convection layer. Central statistical quantities of the flow, such as the mean superadiabatic temperature, temperature and density fluctuations, or turbulent Mach numbers are compared in the form of horizontal plane-time averaged profiles. It is found that the additional material parameter dependence causes systematic quantitative changes of all these quantities, but no qualitative ones. A growing temperature power law exponent $\beta$ also enhances the turbulent momentum transfer in the weak stratification case by 40\%, it reduces the turbulent heat transfer by up to 50\% in the strong stratification case.
\end{abstract}

\maketitle

\section{Introduction}
\label{sec:intro}

Turbulent convection which is driven by buoyancy forces occurs in many natural atmospheric and geophysical flows. \cite{Verma2018} The Rayleigh-B\'{e}nard convection (RBC) case has been considered as one canonical flow setup to study such convection systems.\cite{Kadanoff2001,AGLRMP2009,CJ2012} In this configuration consisting of a plane layer of height $H$, the bottom boundary is kept at a higher uniform temperature compared to the top boundary; a negative temperature gradient is thus imposed along the direction corresponding to gravity acceleration which is the $z$-direction. RBC flows satisfy the Oberbeck-Boussinesq (OB) approximation where a linear density dependence is included for the buoyancy term on the right hand side of the momentum balance only. The flow is incompressible. Both simplifications result in a perfect statistical top-bottom reflection symmetry with respect to the midplane at $z=H/2$. In other words, the mean temperature in the bulk of the layer equals exactly the arithmetic mean of the top and bottom temperature. 

In many practical situations, the OB approximation is violated which results in the breaking of the top-bottom symmetry with respect to the midplane. These flows are commonly termed as non-Oberbeck-Boussinesq (NOB) flows. NOB effects appear due to various reasons: (N1) heights of the convection layer are larger than the corresponding scale heights \cite{SKRMPhys2020}, (N2) additional physics such as phase changes in the atmosphere \cite{stevens2005,pauluis2011,mellado2017}, or (N3) dependencies of material properties, such as dynamic viscosity $\mu(T,p)$ and thermal conductivity $k(T,p)$, on  temperature and pressure \cite{urban2012,couston2017,pandey2021non,pandey2021nona,macek2023} Reason (N1) for a departure from the OB approximations is connected to compressibility effects \cite{frohlich1992,VerWSAJ2015,JPJJSPRF2023,JPJJSJFM2023} as observed for example in giant gas planets \cite{gastine2012effects,jones2009}, planetary mantles \cite{RicardGJI2022} and in solar convection close to the surface \cite{SKRMPhys2020}. 

In order to focus on genuine effects due to compressibility and to disentangle the former effect from reason (N2), it has been common in parts of the literature to assume constant material properties, see e.g. Verhoeven \textit{et al.}\cite{VerWSAJ2015} and Jones \textit{et al.} \cite{jones2022} for comparisons of the fully compressible case with the anleastic limit or Alboussi\`{e}re \textit{et al.} \cite{ACDLRJFM2022} for the limit of very large $Pr$. Panickacheril  and Schumacher \cite{JPJJSJFM2023,JPJJSPRF2023} studied different regimes of compressible convection under constant material properties which generated in parts highly asymmetric flow cases.  NOB effects due to variable material properties in RBC have been studied in direct numerical simulations\cite{horn2011,HornJFM} and in controlled laboratory experiments. \cite{AADGL2007,macek2023} They become relevant when high-Prandtl-number working fluids, such as oils, are used or when the experiments have to be operated close to the critical line in the pressure-temperature plane at very high Rayleigh numbers. Shcheritsa et al.\cite{PLA2008} and Pandey \textit{et al.} \cite{pandey2021non, pandey2021nona} focused on temperature-dependent thermal conductivity at $Pr=1$ and at $Pr\ll 1$, respectively. In both works, it was shown that the variable thermal conductivity (at constant viscosity) resulted already in a significant asymmetry of the mean temperature profile across the turbulent convection layer. Both studies were inspired by the solar convection case.    

In the present work, we combine the non-Boussinesq effects N1 and N3 and study fully compressible turbulent convection with temperature-dependent material properties thus extending our previous works in refs. \cite{JPJJSPRF2023,JPJJSJFM2023}. To this end, we introduce a power law dependence $k(T)\sim T^{\beta}$ and choose $\mu(T)$ such that the Prandtl number $Pr$ remains a constant across the convection layer. This is still a simplification in comparison to astrophysical setups where the Prandtl number itself is a function of height, $Pr=f(z)$, but it includes one additional physical aspect that is common to these flows, a height-dependent superadiabaticity (which will be detailed in the next subsection). To this end, we present 2 series of three-dimensional direct numerical simulations (DNS) at differently strong degrees of stratification and vary the exponent $\beta$ from 0 to 4.0 in case of a dissipation number $D=0.1$ and from 0 to 0.175 for $D=0.8$. The series at $(D,\varepsilon_{\rm global})=(0.1,0.1)$ stands for the weakly stratified case and was denoted to as OB-like compressible convection.\cite{JPJJSPRF2023} The series at $(D,\varepsilon_{\rm global})=(0.8,0.1)$ is strongly stratified convection (SSC).\cite{JPJJSPRF2023} We study the variation of essential turbulence profiles across the layer with respect to the power law exponent $\beta$.

All simulations in this study will be conducted at a moderate Rayleigh number of $Ra \approx10^5$ and at a Prandtl number $Pr=0.7$. This implies that thermal conductivity and dynamic viscosity will vary in the same way across the layer. This is different to previous studies by Pandey \textit{et al.} \cite{pandey2021non} Our work reports the dependence of mean temperature and density profiles on the exponent $\beta$. Emphasis is also given to the specific entropy and the variation of its mean profile with respect to height. We will show that this quantity can be straightforwardly connected to the local stability properties in our convection system. Furthermore, we analyse the turbulent heat transfer as a function of $\beta$.  

The outline of the manuscript is as follows. In Sec. II, we introduce parameters of the compressible convection flow, and discuss the adiabatic and diffusive equilibrium configurations at constant material properties to keep the manuscript self-contained. Section III presents the flow equations and the numerical simulation method. Section IV specifies the height-dependent superadiabaticity for varying material parameters. Subsequent Sec. V present the results. The manuscript ends with a summary and outlook in Sec. VI.

\begin{table*}
\begin{ruledtabular}
\begin{tabular}{cccccccccc}
$Ra_{\rm sa}$ & $D$ & $\beta$  &  $Re$ & $Nu_B$ & $Nu_T$ &  $\varepsilon(0)$   & $\varepsilon(H)$ & $\varepsilon_{\rm global}$ & $R_{\varepsilon}$     \\
\hline
$10^5$ & $0.1$ & $ 0$ & 70.5  & 4.13 & 4.1  & 0.1 & 0.1 &0.1 & 0 \\
$10^5$ & $0.1$ & $ 0.05$ & 70.1  & 4.05 & 4.01  & 0.099 & 0.101 & 0.1 & 0.022 \\
$10^5$ & $0.1$ & $ 0.1$ & 71  & 4.17 & 4.11  & 0.098 & 0.102 & 0.1 & 0.044 \\

$10^5$ & $0.1$ & $ 0.5$ & 74.6  & 4.4 & 4.02  & 0.09 & 0.112 & 0.1 & 0.223 \\
$10^5$ & $0.1$ & $ 1.0$ & 77.3  & 4.8 & 4.01  & 0.08 & 0.125 & 0.1 & 0.449 \\
$10^5$ & $0.1$ & $ 1.5$ & 82  & 5.1 & 3.9  & 0.07 & 0.139 & 0.1 & 0.678 \\

$10^5$ & $0.1$ & $ 3.0$ & 93.6  & 6.8 & 3.8  & 0.04 & 0.188 & 0.1 & 1.403 \\

$10^5$ & $0.1$ & $ 4.0$ & 100.7  & 8.8 &  3.8 & 0.03 & 0.228 & 0.1 & 1.932 \\

\hline 
$10^5$ & $0.8$ & $ 0.0$ & 61  & 2.8 & 2.7  & 0.1 & 0.1 & 0.1 & 0 \\
$10^5$ & $0.8$ & $ 0.05$ & 62  & 3.1 & 1.9  & 0.07 & 0.17 & 0.1 & 1.05 \\
$10^5$ & $0.8$ & $ 0.1$ & 59  & 4.12 & 1.6  & 0.04 & 0.25 & 0.1 & 2.15 \\
$10^5$ & $0.8$ & $ 0.15$ & 57  & 13.6 & 1.4  & 0.01 & 0.34 & 0.1 & 3.31 \\
$10^5$ & $0.8$ & $ 0.175$ & 56  & -- & 1.4  & $\approx$ 0 & 0.37 & 0.1 & 3.67 \\
\end{tabular}
\end{ruledtabular}
\caption{\label{tab:table1} List of parameters and quantities of the two series of direct numerical simulation. The superadiabatic Rayleigh number $Ra_{\rm sa}$ is in all cases an approximate value. Furthermore, we list the dissipation number $D$, the exponent $\beta$, the Reynolds number $Re$, the Nusselt numbers $Nu_B$ and $Nu_B$ at the bottom and top of the layer, the superadiabaticity at the bottom and the top as well as the global value, and the ratio $R_{\varepsilon}$.}
\label{tab1}
\end{table*}

\section{Parameters and equilibria}

An infinitely extended convection layer is governed by two dimensionless parameters, the Rayleigh number $Ra$ and the Prandtl number $Pr$ 
\begin{equation}
Ra = \alpha \Delta T\frac{g H^{3}}{\nu \kappa}\quad\mbox{and}\quad 
Pr = \frac{\nu}{\kappa}\,.  
\end{equation}
Here $\alpha$, $g$, $\Delta T$, $\nu=\mu/\rho$, and $\kappa=k/(C_p\rho)$ are the thermal expansion coefficient, acceleration due to gravity, outer temperature difference across the layer, kinematic viscosity, and thermal diffusivity respectively. The definition of the Rayleigh number is the one that is typically used in the OB case. It has to be adapted for the present configuration as we will see further below.   

Compressibility introduces two further independent dimensionless parameters. The first parameter, which is a measure of the degree of stratification in compressible convection, is the dissipation number $D$ defined as 
\begin{equation}
D = \frac{gH}{C_{p} T_{B}}\,.
\label{eq:one}
\end{equation}
Here $C_{p}$ and $T_{B}$ are the specific heat at constant pressure and the prescribed temperature at the bottom plate, respectively. This corresponds to the temperature gradient of the system when a purely adiabatic process is present. The adiabatic state is obtained from the hydrostatic equilibrium
condition for the pressure field $P$,
\begin{equation}
\frac{d{P_{a}}}{dz}=-g\rho_{a}\,,
\end{equation}
together with the equation of state for an ideal gas ($R$ is the gas constant),
\begin{equation}
 P_{a}=\rho_{a} R T_{a}\,.
 \end{equation}
The subscript \textit{``a"} indicates the adiabatic equilibrium state. We also use the isentropic relations between the thermodynamics variables, $P^{a}/P_{B}=(\rho^{a}/\rho_{B})^{\gamma}$ where $\gamma= C_{p}/C_{v}$ is the ratio of specific heat at constant pressure, $C_{p}$, and volume, $C_{v}$, respectively. The resulting adiabatic equilibrium profiles depend on the vertical coordinate $z$ only and are given by
\begin{align}
T^{a}(z)&=T_{B} \left(1- D \frac{z}{H}\right)\,,\\
\rho^{a}(z)&=\rho_{B} \left(1- D \frac{z}{H}\right)^{m}\,,\\
P^{a}(z)&=P_{B} \left(1- D \frac{z}{H}\right)^{m+1}\,,
\end{align} 
for $0\le z\le H$. In the equations, $m= 1/\left(\gamma - 1 \right)$. Note that the bottom values of all three thermodynamic state variables are the corresponding reference values in the following. Thus the dry adiabatic lapse rate is consequently given by 
\begin{equation}
\frac{dT^{a}}{dz }=-\frac{g}{c_p}.
\end{equation}
One also observes that the dissipation number, $D= gH/C_{p}T_{B}= \left[T^{a}
\left(H\right) - T_{B}\right] /T_{B}$ which defines the temperature drop across the layer during an adiabatic or isentropic process. Furthermore, we underline that the adiabatic equilibrium profile is independent of the variation of the material parameters with temperature or pressure.  

For the linear instability to grow and for fluid motion to occur, the imposed temperature at the top plate must be lower than that corresponding to a pure adiabatic process, $T(H) < T^{a}(H)$. In other words, the actual gradient should exceed the adiabatic one.\cite{jeff1930} Thus, the other compressibility parameter is the superadiabaticity, $\varepsilon_{\rm global}$, defined as 
\begin{equation}
\varepsilon_{\rm global}=  \frac{T^{a}(H) - T(H)}{T_{B}}=\frac{T^{a}(H) - T_T}{T_{B}}\,,  
\label{eq:three}
\end{equation}
where $T_T=T(H)$ is the prescribed temperature at the top boundary. The superadiabaticity $\varepsilon_{\rm global}$ represents the excess temperature gradient from the adiabatic gradient quantified by the parameter $D$. A more general definition for superadiabaticity following ref. \cite{jones2022} can be given as follows,
\begin{equation}
\varepsilon \left( z \right)= -\frac{H}{T_{B}}
\frac{d\overline{T}(z)}{dz} - D\,.
\label{eq:gen_sup}
\end{equation}
Here $\overline{T}(z)$ is the conductive equilibrium profile, i.e., the heat is transported solely by diffusion from the bottom to the top. For a constant thermal conductivity $k$, we get 
\begin{equation}
\overline{T}(z)=-\frac{T_B-\overline{T}(H)}{H} z +T_B\,.
\end{equation}
Plugged into \eqref{eq:gen_sup} leads to \eqref{eq:three}. To conclude, in general superadiabaticity is a function of  depth $z$; for a constant thermal conductivity however,  $\varepsilon(z)$ is a constant and equal to $\varepsilon_{\rm global}$.

\section{Equations and Simulations}

The three-dimensional equations of motion for compressible convection are given by   
\begin{subequations}
\begin{align}
\partial_t \rho  + \partial_i (\rho u_{i}) &= 0
\label{eq:mass}\,,\\
\partial_t (\rho u_{i}) + \partial_j (\rho u_{i} u_{j})  &= -\partial_i p  + \partial_j \sigma_{ij} - \rho g \delta_{i,3}
\label{eq:mom}\,,\\
\partial_t (\rho e) + \partial_j (\rho e u_{j} ) &= -p \partial_i u_{i} +
\partial_i(k \partial_i T) + \sigma_{ij} S_{ij}
\label{eq:ener}\,,\\
p &= \rho R T \quad\textrm{ where }\; R= C_{p} -C_{v}.
\end{align}
\end{subequations}
These equations correspond to mass, momentum and energy conservation laws along with the equation of state of an ideal gas. Quantity $R$ denotes the gas constant. Here, $\rho$, $\rho u_{i}$, $p$, $\rho e$, $T$ are the mass density, momentum density components, pressure, internal energy density, and temperature, respectively. The viscous stress tensor depends now on temperature and is given by 
\begin{equation}
\sigma_{ij}=2\mu(T)S_{ij}-\frac{2}{3}\mu(T)\delta_{ij}(\partial_k u_k)
\end{equation}
with the rate of strain tensor $S_{ij} = (\partial_i u_j+\partial_j u_i)/2$.  

We will assume a power law with respect to temperature $T$ for the dynamic viscosity, $\mu(T)$ in our simulations. The thermal conductivity $k(T)$ is related to the viscosity through the Prandtl number, 
\begin{equation}
k(T)= \frac{\mu(T) C_{p}}{Pr}\,.    
\end{equation}
In the present work, we fix $Pr= 0.7$. $C_{p}$ and $C_{v}$ correspond to specific heat at constant pressure and volume, respectively. Their ratio, $\gamma= C_{p}/C_{v}= 1.4$ for a diatomic gas. The specific internal energy is defined as $e=C_{v}T$. In the next section, we will need a different version of the energy equation \eqref{eq:ener} for the discussion of the turbulent heat transfer. It is based on the temperature field $T$ and given by
\begin{align}
C_p\partial_t(\rho T)&+C_p\partial_j(\rho u_j T)-\partial_t p -u_j\partial_j p 
\nonumber\\
&=\partial_j(k(T)\partial_j T)+\sigma_{ij}(T)S_{ij}\,.
\label{eq:T}
\end{align}

A uniform grid is used in $x$-- and $y$--directions along with periodic boundary conditions. In wall-normal $z$--direction, a non-uniform grid with a point clustering near the walls is taken, which follows a hyperbolic tangent stretching function. Spatial derivatives are calculated by a 6th-order compact  scheme  for all points except near the walls \citep{lele1992,BDBjfm2022}; there 4th- and 3rd--order compact schemes are used at the last two grid points near the wall. No-slip, isothermal boundary conditions are applied at the top and bottom.  The boundary condition for $p$ is evaluated using the $z$-component of the momentum equation at $z=0, H$, $\partial p/\partial z =  \partial \sigma_{iz}/\partial x_{i} - \rho g$. The fields are advanced in time by a low storage 3rd-order Runge-Kutta with a Courant number of ${\rm CFL}=0.5$. All simulations are carried out in a Cartesian slab with quadratic cross section and an aspect ratio of $\Gamma=L/H=4$. 

Some important parameters are summarized in table \ref{tab1}. This includes the global measures of turbulent heat and momentum transfer, the Nusselt number $Nu$ and the Reynolds number $Re$. The Reynolds number $Re$ is defined as follows,
\begin{equation}
    Re = \frac{\langle \rho \rangle_{V,t} \langle u_i^2\rangle_{V,t}^{1/2} H}{\langle \mu(T)\rangle_{V,t} }\,,
\end{equation}
since the dynamic viscosity will be temperature-dependent, $\mu(T)$.  The Reynolds number for different to be introduced later are given in Table. 1.

Also differently to ref. \cite{JPJJSJFM2023}, the Nusselt number $Nu$, which relates the total heat transfer to the diffusive one, will not be constant across the layer. If the turbulence is in a statistically stationary regime and the eq. \eqref{eq:T} is averaged over the homogeneous horizontal directions then one obtains the following $z$-dependent mean balance
\begin{align}
C_p\langle\rho u_j T\rangle_{A,t}&-\int_0^z \langle u_j\partial_j p\rangle_{A,t} dz^{\prime} -\int_0^z \langle \sigma_{ij}(T)S_{ij}\rangle_{A,t} dz^{\prime}
\nonumber\\
&-\Bigg\langle k(T) \frac{dT_{\rm sa}}{dz}\Bigg\rangle_{A,t} = \mbox{const.} 
\label{eq:T1}
\end{align}
The last term on the left hand side of \eqref{eq:T1} is used to define the Nusselt number at the bottom and top plates which are given by 
\begin{align}
    Nu_B = - \frac{H}{\varepsilon(0)T_B} \Bigg\langle \frac{d T_{\rm sa}}{dz}\Bigg|_{z=0}\Bigg\rangle_{A,t}
    \label{Nu_bot}
\end{align}
and 
\begin{align}
    Nu_T = - \frac{k(T_T) H}{\varepsilon(H) k_0 T_B} \Bigg\langle \frac{d T_{\rm sa}}{dz}\Bigg|_{z=H}\Bigg\rangle_{A,t}\,,
    \label{Nu_top}
\end{align}
respectively. The details on the constant reference thermal conductivity $k_0$ and the height-dependent superadiabaticity will follow now in Sec. IV. Here, the superadiabatic temperature field is given by 
\begin{equation}
T_{\rm sa}({\bm x},t)=T({\bm x},t)-T_a(z)\,.
\label{eq:T_sa}
\end{equation}
Equations \eqref{Nu_bot} and \eqref{Nu_top} will give different magnitudes for the present cases with variable material parameters and are listed in Table. 1. In case of temperature-dependent thermal conductivity, which we will discuss in the next section in detail, it will be shown that the superadiabaticity for the strongest material property variations goes to zero, $\varepsilon \rightarrow 0$. Thus, the Nusselt number $Nu_B$ in \eqref{Nu_bot} is not defined for this case.  However,  $\varepsilon\left(H\right)$ is always finite and we can define $Nu_T$ by \eqref{Nu_top} for all cases. Consequently, we will use the Nusselt number at the top $Nu_T$ to quantify the turbulent heat transfer efficiency.

\section{Variable Thermal Conductivity and Superadiabaticity  }
  
In solar convection, the thermal conductivity follows to $k(T) \sim T^{3}$ since heat is carried by photons to the outside. Inspired by such a scaling behavior, we assume here that the thermal conductivity $k$ takes the following general form across the convection layer,
\begin{subequations}       
\begin{equation}
    k(T)  = k_0 \left[\frac{T}{T_{B}}\right] ^{\beta}\,, 
\end{equation}
with the constant $k_0$. In order to obtain a constant Prandtl number $Pr$ across the thermal convection layer (as already discussed in the introduction), we assume that the dynamic viscosity $\mu$ varies as 
\begin{equation}
    \mu(T)  = Pr \frac{k_0}{C_{p}} \left[\frac{T}{T_{B}}\right] ^{\beta}. 
\end{equation}
\end{subequations}
The resulting conductive temperature profile, which is a function of the vertical $z$--coordinate only, is given by  
\begin{equation}
\left(\frac{\overline{T} \left(z\right) } {T_{B}} \right)^{\beta + 1} = 1 + \left[ \left(\frac{T_{T}}{T_{B}}\right)^{\beta +1}  - 1    \right]\frac{z}{H}\,.
\label{prof}
\end{equation}
Here, $T_{T}$ is the temperature at the top plate, i.e., $T\left(H\right)$. From the general power law profile \eqref{prof} the following expression for $\varepsilon\left( z \right)$ in eq.~(\ref{eq:gen_sup}) can be derived 
\begin{equation}
\varepsilon \left( z \right)= \left[\frac{T_{B}^{\beta}}{\beta + 1}\right] \left[\frac{1}{\overline{T}^{\beta}\left(z\right)}\right]\left[1 -  \left(\frac{T_{T}}{T_{B}}\right)^{\beta +1}      \right] - D\,.
\label{eq:eps_z}
\end{equation} 
Thus, the superadiabaticity at the bottom ($z=0$) and top ($z=H$) plates are given by the following relations
\begin{align}
        \varepsilon(0) = &\left[\frac{1}{\beta +1}\right]\left[ 1 - (1-\varepsilon_{\rm global} - D)^{\beta +1}     \right] - D
\label{eq:eps_z0}\,,\\
        \varepsilon(H) = &\left[\frac{1}{\beta +1}\right] \left[\frac{1}{1 - \varepsilon_{\rm global} - D}\right]^{\beta} \times\nonumber\\  &  \left[ 1 - (1 - \varepsilon_{\rm global} - D)^{\beta+1}     \right] - D\,.
\label{eq:eps_zH}
\end{align}
Note that the ratio of bottom to top temperature is written in terms of global superadiabaticity, $\varepsilon_{\rm global}$ and dissipation number, $D$. Thus $T_{T}/T_{B} =  1 - \varepsilon_{\rm global} - D$.
\begin{figure}
\includegraphics[width=1.0\linewidth]{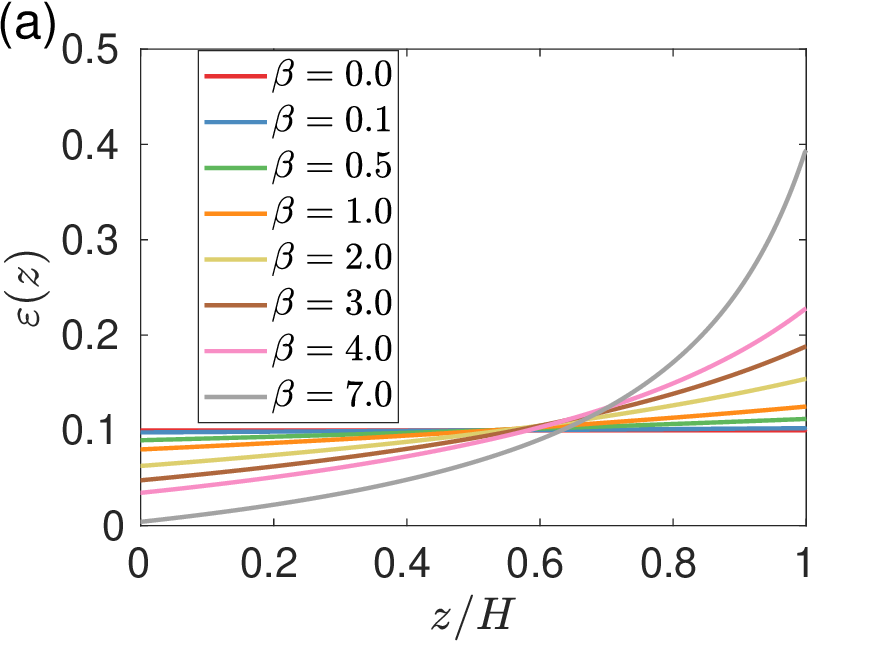}
\includegraphics[width=1.0\linewidth]{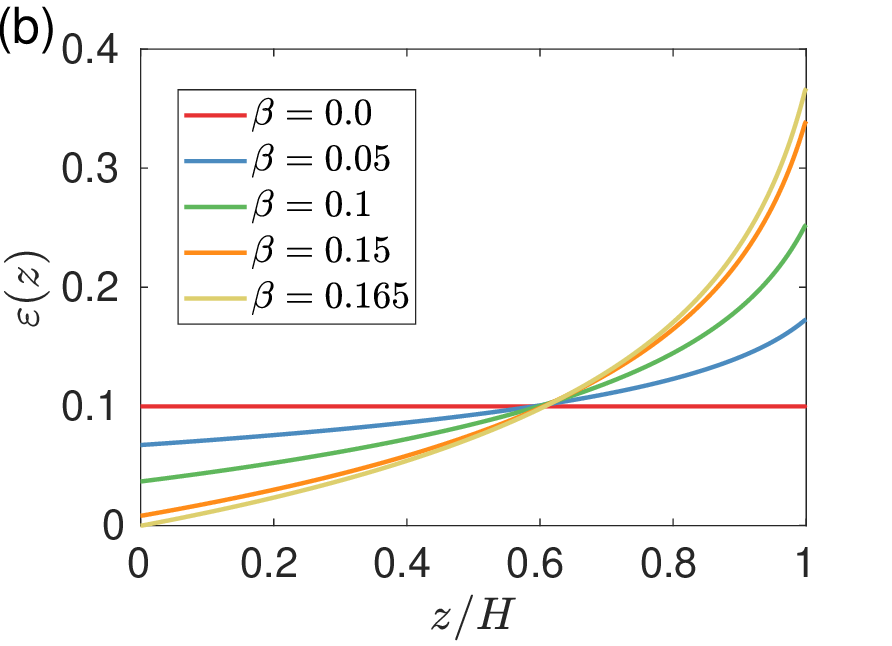}
\caption{\label{fig:eps_z}  Variation of superadiabaticity $\varepsilon(z)$ across depth $z/H$ as function of power law exponent $\beta$. The exponents of the corresponding superadiabaticity profiles are given in the legends of both panels. (a) Series for dissipation number $D=0.1$ and superadiabaticty $\varepsilon_{global}= 0.1$. (b) Series for $D=0.8$ and $\varepsilon_{global}= 0.1$. In the latter case the power law exponent could not be varied so strongly as in the first series.}
\end{figure}
\begin{figure}
\includegraphics[width=1.0\linewidth]{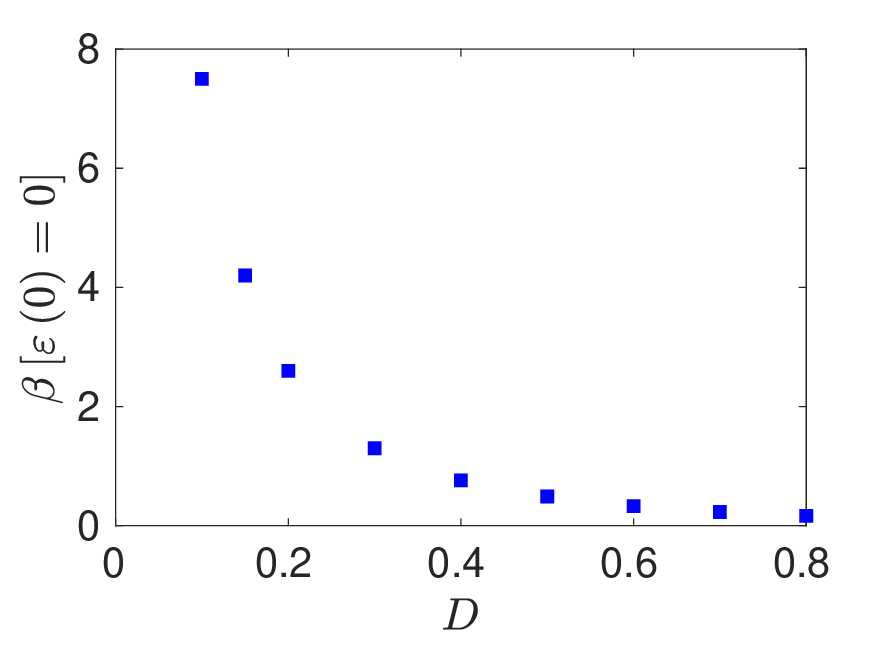}
\caption{\label{fig:beta_D}  The upper limit of the power law exponent $\beta$ as a function of dissipation number, $D$. For $D=0.8$, one gets $\beta\le 0.165$.}
\end{figure}

Comparing eqns.~(\ref{eq:eps_z})--(\ref{eq:eps_zH}), one finds that $\varepsilon(z)$ increases with height $z$ for a power law with an exponent $\beta > 0$. For constant thermal conductivity, i.e., for $\beta=0$, the superadiabaticity simplifies to $\varepsilon(z) = \varepsilon_{\rm global}$ everywhere. Here, we obtain $\varepsilon_{\rm global}$ by averaging with respect to the depth of the layer, 
\begin{equation}
   \varepsilon_{\rm global} = \frac{1}{H}\int_{0}^{H} \varepsilon\left(z\right) dz\,.
\end{equation}
Note that the profiles are chosen such that $\varepsilon_{\rm global}=0.1$ for all cases, see also table I. This guarantees the consistent comparison with the runs at constant material properties which were conducted in our previous studies.\cite{JPJJSJFM2023,JPJJSPRF2023}

One can also estimate the degree of departure from the global superadiabaticity across the domain by introducing the following new parameter (which we also list in table I)
\begin{equation}
R_{\varepsilon}= \frac{\varepsilon\left(H\right) -\varepsilon\left(0\right) }{\varepsilon_{\rm global}}
\end{equation}
The global (superadiabatic) Rayleigh number is defined as follows
\begin{subequations}
\begin{equation}
    Ra_{\rm sa} = \frac{\langle \rho \rangle^{2}_{V,t} g \varepsilon_{global} H^{3}}{\mu\left(0\right) k\left(0\right)}\,. 
\end{equation}
However in general, the (superadiabatic) Rayleigh number varies with layer depth $z$ as 
\begin{equation}
    Ra_{\rm sa}(z) = \frac{\langle \rho \left(z\right)  \rangle^{2}_{A,t} g \varepsilon \left(z\right) H^{3}}{\mu\left(z\right) k\left(z\right)}, 
    \label{eq:varra}
\end{equation}
\end{subequations}

At general compressibility conditions with variable thermal conductivity, the superadiabaticity is strong function of depth as evident from  Figs.~\ref{fig:eps_z} (a,b).  For both cases, the difference between the top and bottom increases with the exponent $\beta$. Moreover, although $\varepsilon_{\rm global}= 0.1$ in all simulation runs, there are regions in the layer --particularly near the top boundary-- where $\varepsilon(z) > 0.1$. Here, an anelastic approximation, which requires $\varepsilon_{\rm global}\ll 1$, may not be valid anymore.  Our recent work with constant material parameters \cite{JPJJSJFM2023} showed that at $Ra \approx 10^{6}$ and for very strong stratification with $D \gtrsim 0.65$ the anelastic approximation is definitely not valid, even for $\varepsilon_{\rm global} =0.1$. In the present case, we extend these studies to stratified compressible convection with variable material properties which lead to  asymmetry in superadiabaticity across the domain and thus amplify the asymmetry between top and bottom of the constant material parameter case even further. 

This effect of an asymmetry of the superadiabaticity across the convection layer is absent in the incompressible Oberbeck-Boussinesq RBC case, thus our current study is also different from the previous works which considered variable material property effects.\cite{pandey2021non, pandey2021nona}. A quick  calculation shows that under OB conditions $ \left(\varepsilon_{\rm global} \lesssim 10^{-3} \textrm{  and }   D \lesssim 10^{-3} \right)$, even with an exponent of $\beta= 10$ would  only result in $\varepsilon(0)=0.00098$ and $\varepsilon(H)=0.001$. These are  approximately uniform and thus equal to $\varepsilon_{\rm global}$. Consequently, for OB convection the Rayleigh number is practically uniform despite strong power law behavior of dynamic viscosity and thermal conductivity. This is of course due to the implicit  assumption in OB convection that $(T_{B} - T_{T})/T_B  \ll 1$. However, for fully compressible convection where either $\varepsilon \not \ll 1$ or $D \not \ll 1$, the  Rayleigh number changes across the depth due to both, the variation in the density as well as the variation in the material properties, as seen in eq. \eqref{eq:varra}.       

\begin{figure}
\includegraphics[width=1.0\linewidth]{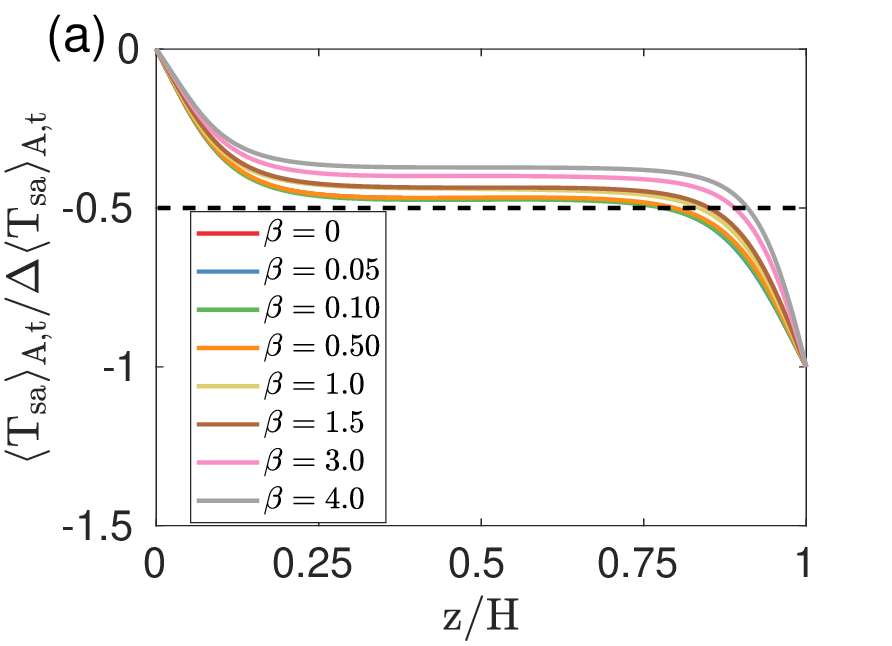}
\includegraphics[width=1.0\linewidth]{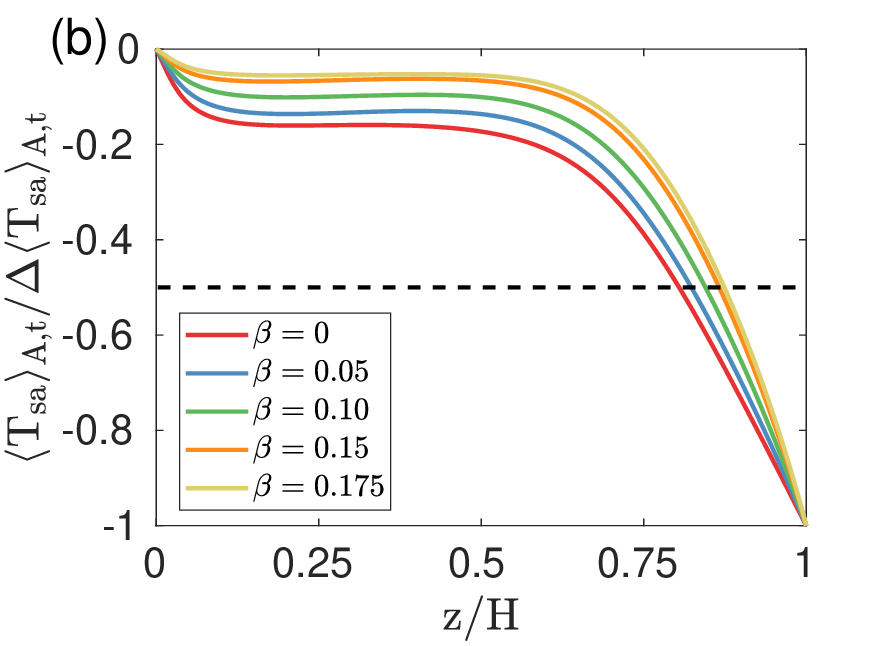}
\caption{\label{fig:Tsa} Superadiabatic temperature profiles obtained as averages over the horizontal cross section and time. (a) $D= 0.1$ and (b) $D= 0.8$. The corresponding values of $\beta$ are indicated in the corresponding legends. The data are normalized by the total superadiabatic temperature difference such that they are comparable.}
\end{figure}

From Figs.~\ref{fig:eps_z} (a,b) and eq.~(\ref{eq:eps_z0}), we can infer that beyond $\beta= 7.5$ and 0.165, the superadiabaticity at the bottom plate becomes negative for $D=0.1$ and $D= 0.8$, respectively. This holds at $\varepsilon_{\rm global}= 0.1$. Physically, a variable thermal conductivity  can introduce stratification near the bottom boundary. Thus, there is an  upper limit for the exponent $\beta$ for each specific dissipation number $D$ at each $\varepsilon_{\rm global}= 0.1$  when $\varepsilon\left(0\right)= 0$. This upper limit of $\beta$ is shown as a function of the dissipation number $D$ in Fig.~\ref{fig:beta_D}. 

In this manuscript, we will consider two series of direct numerical simulation runs, one at $D= 0.1$ and a second one at $D=0.8$. For $D=0.1$, we have $\beta$ values between  0 to 4, whereas for $D= 0.8$ they will vary from 0 to 0.175. It should be noted that for $D= 0.1$, even our highest exponent $\beta = 4.0$ is well below the upper limit of 7.5 (see paragraph above) and thus $\varepsilon > 0$. However, for $D= 0.8$, our largest exponent $\beta= 0.175$ which is close to the limit and thus $\varepsilon\left(0\right) \approx 0$.     

\begin{figure}
\includegraphics[width=1.0\linewidth]{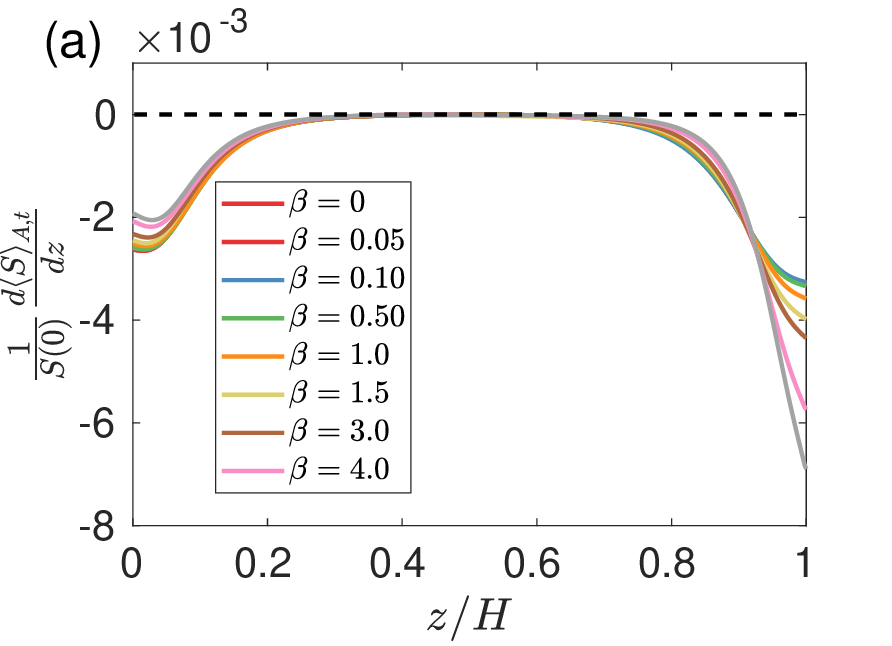}
\includegraphics[width=1.0\linewidth]{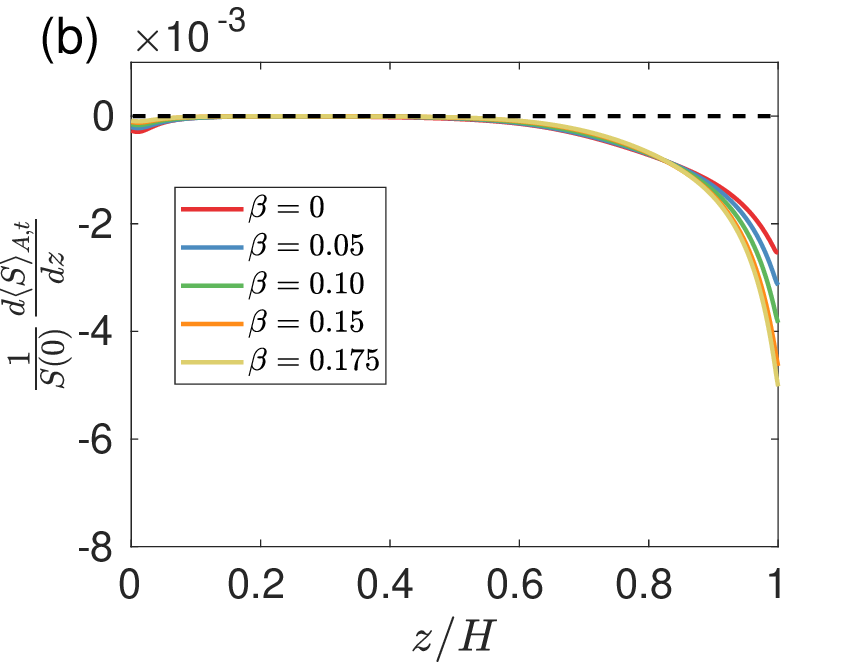}
\caption{\label{fig:tot_entdiss} Variation of the normalized specific entropy $S$ across depth $z$ for (a) $D= 0.1$ and (b) $D= 0.8$.}
\end{figure}
\begin{figure}
\includegraphics[width=1.0\linewidth]{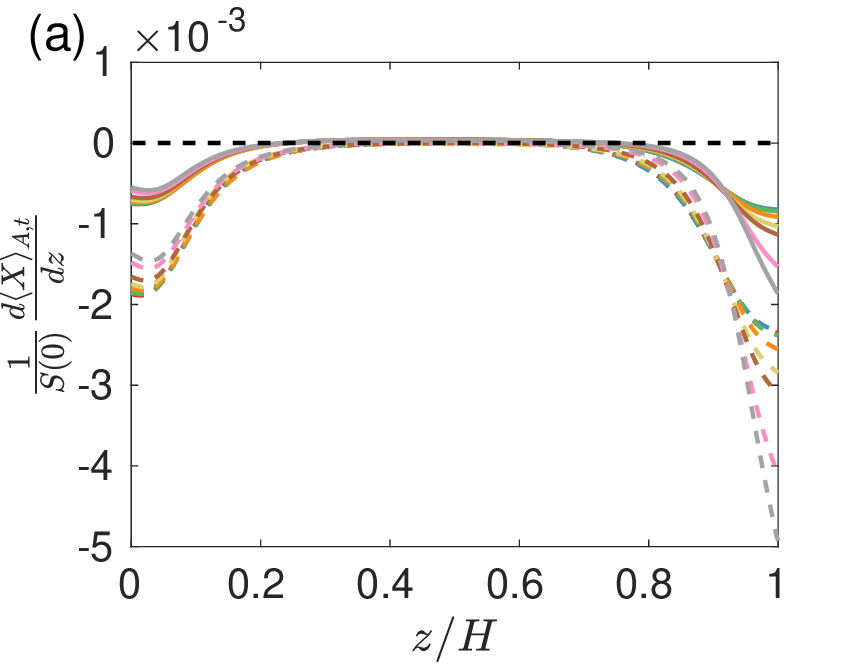}
\includegraphics[width=1.0\linewidth]{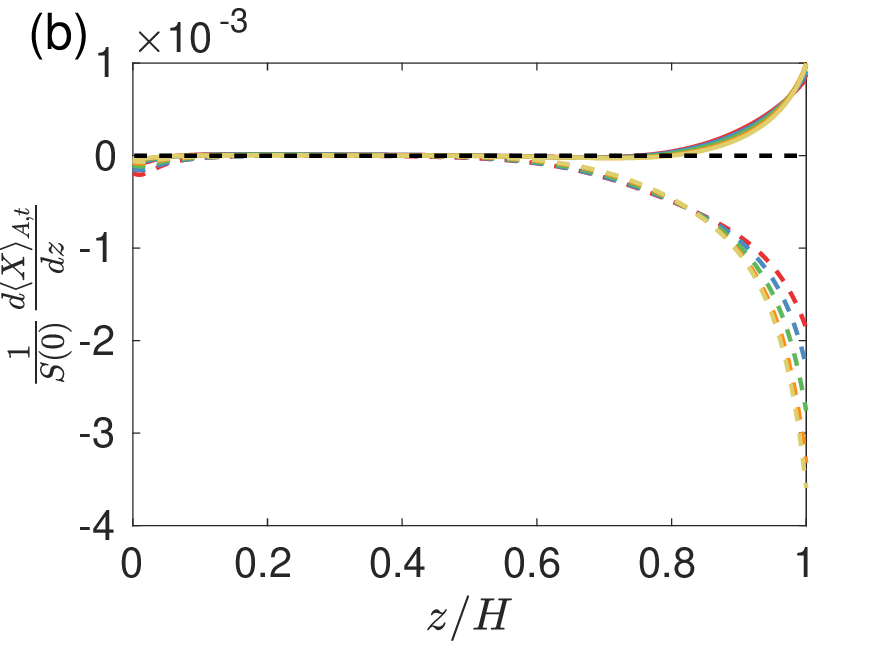}
\caption{\label{fig:ent_cont} Relative contributions from the mean superadiabatic density and temperature variations with respect to $z$ to the variation of the mean specific entropy. (a) $D= 0.1$  and (b) $D= 0.8$. For the dashed lines, $d\langle X \rangle_{A,t}/dz = (C_{v}/\langle  T \rangle_{A,t})\, d\langle  T_{\rm sa} \rangle_{A,t}/dz$. For the solid lines, we have $d\langle X \rangle_{A,t}/dz =-(R/\langle  \rho \rangle_{A,t})\, d \langle  \rho_{\rm sa} \rangle_{A,t}/dz$ where $\rho_{\rm sa}$ is the superadiabatic mass density. The color coding for $\beta$ in both panels is exactly the same as in Fig. \ref{fig:tot_entdiss}.}
\end{figure}

\section{Results}
  
\subsection{Superadiabatic temperature profiles} 
One of the main effects of variable material properties is the imposed variable superadiabaticity across the depth as we have seen in the last section. In Fig. \ref{fig:Tsa}, we plot the  normalized mean superadiabatic temperature for $D= 0.1$ in panel (a) and  $D= 0.8$ in panel (b) for various exponents $\beta$. The quantity is defined by eq. \eqref{eq:T_sa} and $\langle\cdot\rangle_{A,t}$ denotes a combined horizontal plane-time average. Indeed, one observes in the figure that for all the cases the bulk regions have a constant superadiabatic temperature caused by the turbulent mixing of the temperature field.  As expected, the temperature drop at the top boundary is greater than at the bottom boundary for the strongly stratified case of $D= 0.8$ in comparison to the weakly stratified one at $D= 0.1$. The stronger the degree of stratification, the larger is the offset from the symmetric bulk value of $\langle T_{\rm sa}\rangle_{A,t}/\Delta \langle T_{\rm sa}\rangle_{A,t}=0.5$. Here, we also observe that this asymmetry is further enhanced when the variation of the material properties across the layer increases. This is seen for both series of simulation data, even though the range of accessible $\beta$ differs. As $\beta$ increases, the bulk temperature comes ever closer to the temperature value of the bottom boundary, thus implying a larger drop across the top boundary.

By construction, the value of dynamic viscosity and  thermal conductivity are at the same ratio at the bottom boundary for every $\beta$. Thus for $D=0.1$, we observe that the temperature profiles practically collapse near the bottom. Even though the exponent $\beta$ varies between 0 and 4, the difference between the profiles remains moderate. The sensitivity with respect to $\beta$ is strongly enhanced for the strongly stratified case at $D= 0.8$; the departure of the profiles for the different exponents $\beta$ in comparison to the case of $\beta= 0$ is clearly observable, even near the bottom boundary.  As $\beta$ is increased, the thermal conductivity $k$ also decreases near the top boundary; thus we observe indeed thinner boundary layers for both series as $\beta$ is increased.

\subsection{Mean specific entropy variation across the layer} 
The superadiabaticity is directly related to the relative change in the entropy across the depth. The variation of the mean specific entropy with respect to layer depth allows us to summarize the effects of both, the height variations of superadiabatic mean temperature and superadiabatic mean density in the quantitative ratio to each other. Figures~\ref{fig:tot_entdiss}(a,b) display the relative change of the mean specific entropy for both dissipation numbers, $D= 0.1$ and $D= 0.8$. A horizontal plane-time average is calculated again. Therefore, we take the vertical derivative of the specific entropy $S$ which is given by 
\begin{subequations}     
\begin{equation}
\frac{d\langle S \rangle_{A,t}}{dz}  = \left[\frac{C_{v}}{\langle T \rangle_{A,t}} \frac{d \langle T \rangle_{A,t}}{dz} - \frac{R}{\langle \rho \rangle_{A,t}} \frac{d \langle \rho \rangle_{A,t}}{dz}\right]\,,  
\label{eq:ent_def}
\end{equation} 
and normalize by the specific entropy value at the bottom, $S(0)$, which is defined as
\begin{equation}
S(0) = C_{v} \log_e \langle T \rangle_{A,t}(0) - R \log_e \langle \rho \rangle_{A,t}(0)\,.
\label{eq:ent_bot}
\end{equation}
\end{subequations}
\begin{figure}
\includegraphics[width=1.0\linewidth]{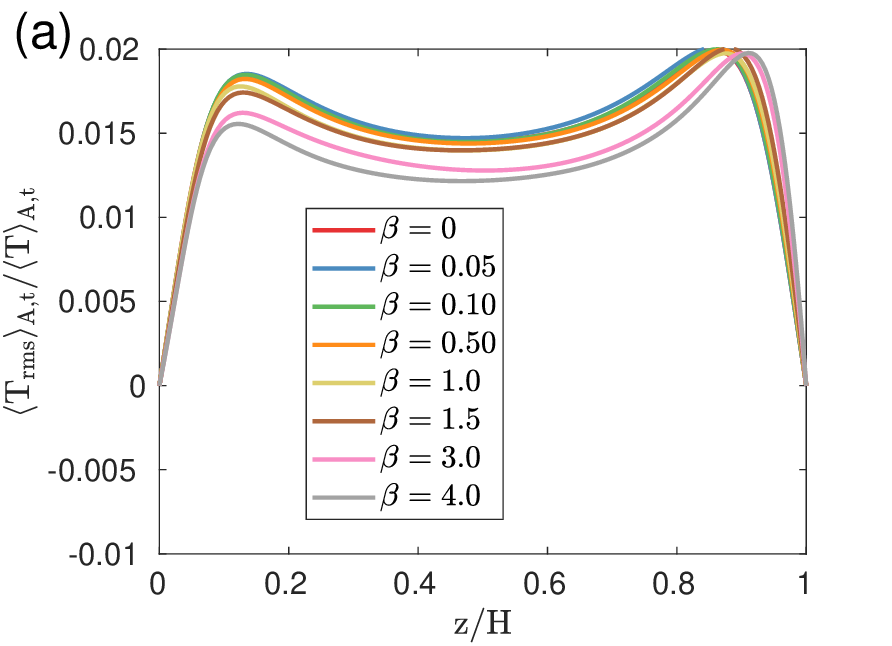}
\includegraphics[width=1.0\linewidth]{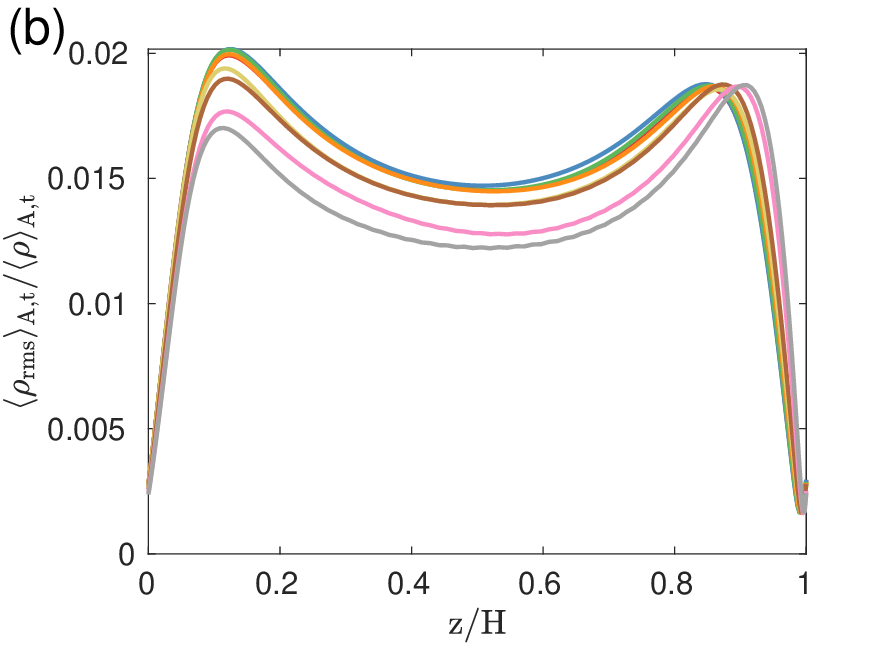}
\caption{\label{fig:temp1} Normalized (a) temperature and (b) density fluctuations for the weakly stratified case at $D= 0.1$. Color coding holds for both panels.}
\end{figure}
\begin{figure}
\includegraphics[width=1.0\linewidth]{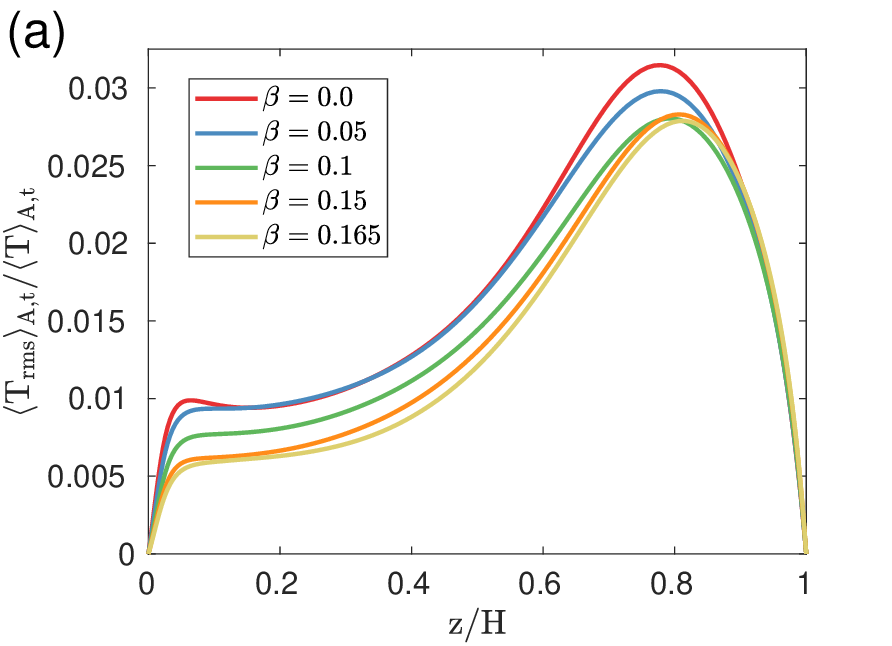}
\includegraphics[width=1.0\linewidth]{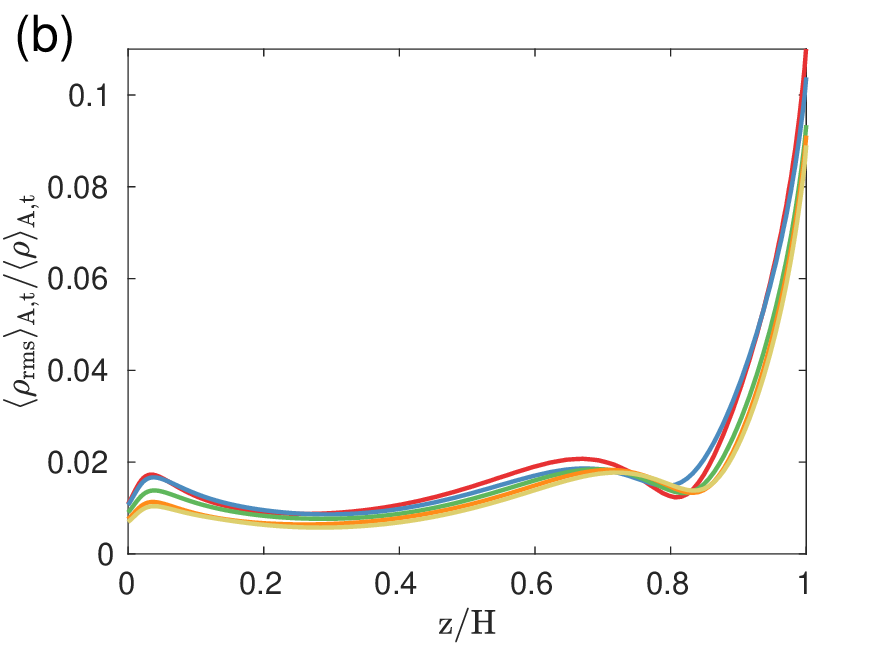}
\caption{\label{fig:temp2} Normalized (a) temperature and (b) density fluctuations for the strongly stratified case at $D= 0.8$. Color coding holds for both panels.}
\end{figure}
The superadiabatic density is defined as
\begin{equation}
\rho_{\rm sa}({\bm x},t)=\rho({\bm x},t)-\rho_a(z)\,.
\label{eq:T_sa}
\end{equation}
From  Fig.~\ref{fig:tot_entdiss}(a), we obtain asymmetric profiles of the mean specific entropy change even for the low dissipation number $D= 0.1$ and all $\beta$. Similar observations can be made for $D= 0.8$ even though the asymmetry due to the strong stratification will dominate over that due to variable property effects. For the high dissipation number $D= 0.8$, it is obvious that the drop near the bottom boundary is negligible compared to that near the top boundary. Overall, it is found that relative change exists, is $\beta$-dependent, but remains small in magnitude.  

From eq. \eqref{eq:ent_def}, it is also clear that the contribution to entropy differences comes from both, the mean superadiabatic temperature and density gradients. Thus we plot the individual contributions of temperature and density derivatives for both $D= 0.1$ and $D= 0.8$ in Fig. \ref{fig:ent_cont}. Note also that we consider the superadiabatic part in the numerator since the adiabatic contributions could cancel each other to give a net zero contribution to the specific entropy gradient.   

In panel (a) of Fig. \ref{fig:ent_cont} for $D= 0.1$, we find that the profiles corresponding to vertical temperature and density derivatives are qualitatively similar. This behavior is similar to that one would expect in an OB approximation where both  temperature and density are negatively correlated, see also \eqref{eq:ent_def}. The additional variable material properties do not fundamentally change this basic correlation between density and temperature for the case of weak stratification at $D= 0.1$.

In panel (b) of the same figure, we plot the contributions from density and temperature to the vertical variation of the mean specific entropy for $D= 0.8$ for all exponents $\beta$. Consistent with our recent study\cite{JPJJSJFM2023}, a change in the behavior at $D= 0.65$ is reflected here in the different signs of mean superadiabatic temperature and density contributions. It is seen that they are no longer negatively correlated near the top boundary. This agrees with positive temperature--density correlations corresponding to an anti-convective, stabilizing behavior. The latter was reported in Panickacheril John and Schumacher.\cite{JPJJSJFM2023} In conclusion, we find that for both, small and large dissipation number $D$, the basic behavior of $d\langle S\rangle_{A,t}/dz$ remains the same. This holds even for strong material property variations. However, non-negligible and systematically increasing changes for growing exponent $\beta$ are obvious for both series at $D= 0.1$ and $D= 0.8$.

\subsection{Density and temperature fluctuation profiles}
In RBC, the strength of the temperature fluctuations plays a significant role for the turbulent heat transport. Exactly these correlations are directly connected to the permanently detaching plumes. Figures \ref{fig:temp1}(a,b) display the plane-time averaged profiles of the root-mean-square values of temperature and density normalized by its own respective mean for the weakly stratified case at $D= 0.1$ and for all $\beta$. Once again, we find that the behavior and magnitude  of density and temperature fluctuations are more or less similar.  The magnitude of both, density and temperature fluctuation profiles in proximity to the bottom wall are the same for all runs and different $\beta$. This is because, the construction of the material dependencies for $k(T)$ and $\mu(T)$ are the same at the bottom boundary for all cases. It will give rise to similar boundary layer thicknesses. However, the maxima of both, density and  temperature fluctuations near the bottom boundary, decrease with growing $\beta$ or stronger material property variations.  This can be directly attributed to the decrease in superadiabaticity, $\varepsilon(z)$ near the bottom boundary which is obtained for increasing $\beta$. 

For the top boundary, we observe that the boundary layer thickness decreases with increasing $\beta$. This is consistent with the fact that as we increase $\beta$, both dynamic viscosity and thermal conductivity also decrease resulting in thinner boundary layer thicknesses. However, despite the fact that the superadiabaticity $\varepsilon(z)$ near the top boundary increases with $\beta$,  the maxima  of both, density and temperature fluctuations, are more or less similar in magnitude for all $\beta$ cases.      

Figures \ref{fig:temp2}(a,b) provide the normalized temperature and density fluctuation profiles across the convection layer for all $\beta$ of the strongly stratified case at a dissipation number of $D= 0.8$. As expected, under strong background stratification, the behavior of temperature and density is no longer qualitatively similar and thus needs to be discussed separately now.  

In panel (a) of the figure for the normalized temperature fluctuation profiles, we observe an asymmetry between the top and bottom boundaries due to strong stratification. The behavior near the bottom boundary is similar to that of the weakly stratified case $D=0.1$. Similarly to the former case, the local fluctuation maximum decreases with increasing $\beta$ (due to decreased superadiabaticity). The profiles are different at the top boundary in comparison to the weakly stratified case. All profiles collapse and lead to similar relative boundary layer fluctuation magnitudes near the top boundary. Again, the local maxima decrease with increasing $\beta$. This behavior seems counter-intuitive as one expects to see increased fluctuations with increased superadiabaticity, see further below for a discussion.  

As expected in panel (b) of the same figure, the structure of the density fluctuation profiles is very different from that of the temperature in the high stratification case at $D= 0.8$. Similar to temperature fluctuations in panel (a), the maxima of density fluctuations at the top boundary at $z= H$ decrease with increasing $\beta$. The bottom boundary behavior is similar to that of $D= 0.1$. This is not surprising as we expect the strongest effects of compressibility and stratification near the top boundary.\cite{JPJJSJFM2023} 

\begin{figure}
\includegraphics[width=1.0\linewidth]{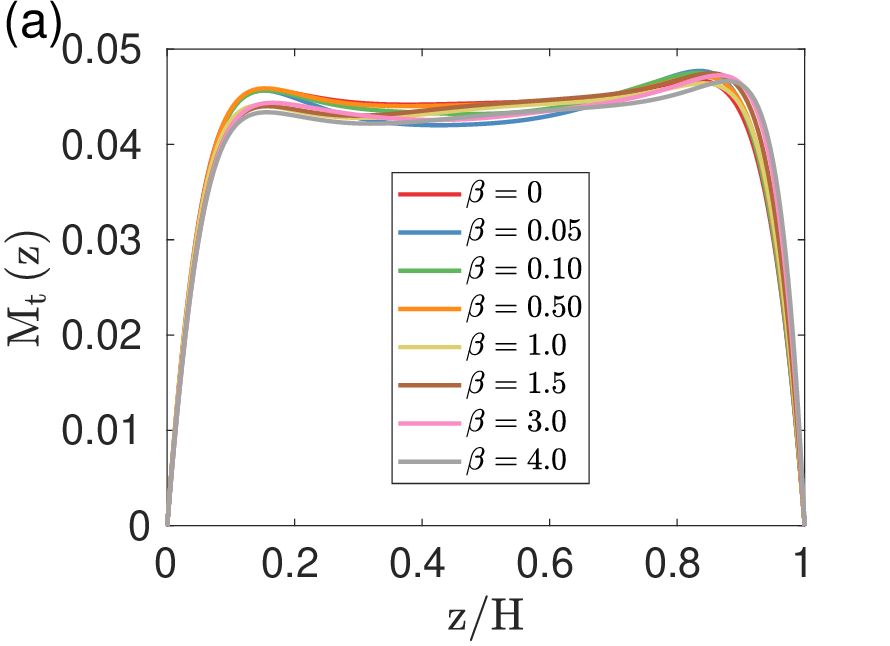}
\includegraphics[width=1.0\linewidth]{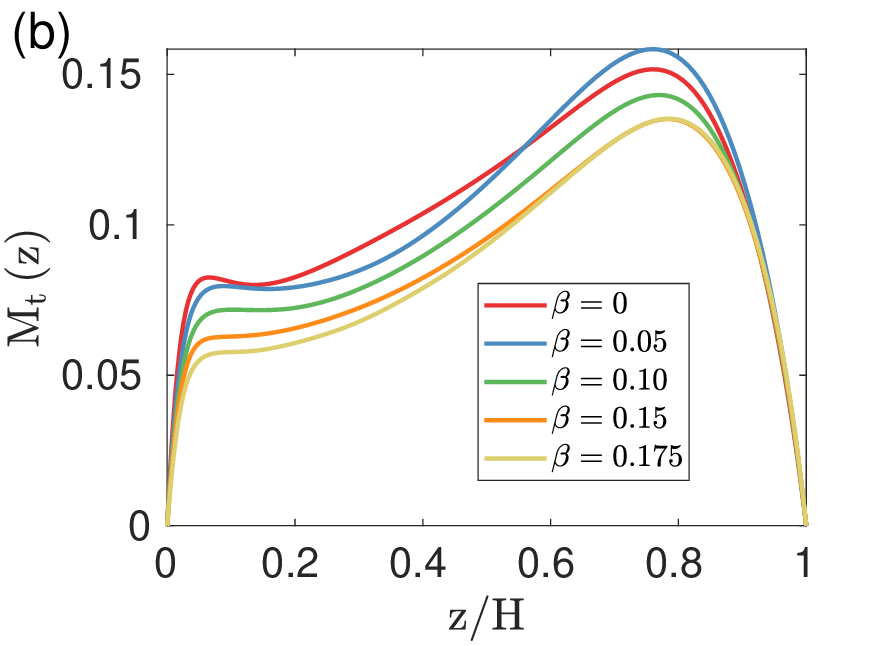}
\caption{\label{fig:mach} Variation of the plane-time averaged turbulent Mach number $M_t$ with depth $z$ for (a) $D= 0.1$ and (b) $D= 0.8$. The color legends indicate the corresponding values of the power law exponent $\beta$.}
\end{figure}

The counter-intuitive behavior near the top boundary, which we stated above, can be rationalized as follows.  Before, let us recall that a completely different dynamics exits here when comparing $D=0.1$ and $D=0.8$ at the same $\varepsilon_{\rm global}$, as analysed in detail in our previous study.\cite{JPJJSJFM2023} For $D<0.65$ normalized mean density and temperature fluctuation profiles are found to collapse; for $D>0.65$ normalized mean pressure and density fluctuation profiles are found to be close. The weak stratification case would thus be similar to OB RBC (where density variations depend linearly on temperature deviations), whereas for the strong stratification case we have a different correlation. As a consequence, slender plume structures fall from the top boundary deep into the bulk and dominate the heat transport.

In the convection setup which is considered here, the stratification is in line with a negative background density gradient with respect to height. A density perturbations should thus be high enough such that this negative density gradient can be overturned locally. Such a process results in a slender plume emanating from the top layer. The picture also implies that under strong stratification, there has to be a minimum level of density fluctuations which can result in the formation of a downward falling plume. Large superadiabaticity near the top would however be in line with a weaker local background stratification; thus lower-magnitude density perturbations can already induce downward falling plumes. In other words, with increased superadiabaticity near the top, which is caused by an increasing exponent $\beta$, the minimum threshold of density perturbations for triggering a local turnover of the background density is decreased. Thus, we observe that the magnitude of fluctuations of density and temperature decrease with $\beta$ for the high stratification case at $D= 0.8$.        

\subsection{Impact on the degree of compressibility} 
Figures \ref{fig:mach}(a,b) display the mean turbulent Mach number, which is given by 
\begin{equation}
M_t(z)=\sqrt{\frac{\langle u_i^2(z)\rangle_{A,t}}{\gamma R \langle T(z)\rangle_{A,t}}}\,,
\end{equation}
as a function of depth $z$ for both $D$. In panel (a) of the figure we observe that variable material properties do not introduce any significant asymmetry for the strength of compressibility effects which is typically monitored by $M_t$. The asymmetry between the top and bottom regions remains small despite the strong background asymmetry in superadiabaticity $\varepsilon(z)$ for the low stratification case at $D= 0.1$. For the series at $D= 0.8$ in panel (b) of the figure, the variable material properties do not introduce an additional asymmetry (as the profiles remain nearly parallel to each other). However the overall strength of the compressibilty is decreased with increasing $\beta$. This finding is also consistent with the reduced thermodynamic fluctuations that we observed near the top boundary before. 

\subsection{Impact on global momentum and heat transfer}
\begin{figure}
\includegraphics[width=1.0\linewidth]{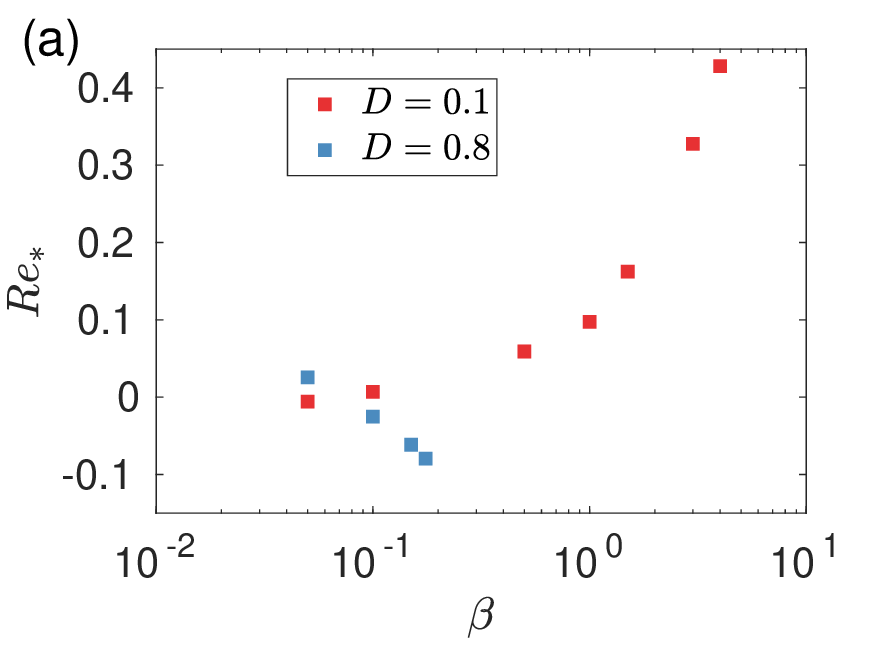}
\includegraphics[width=1.0\linewidth]{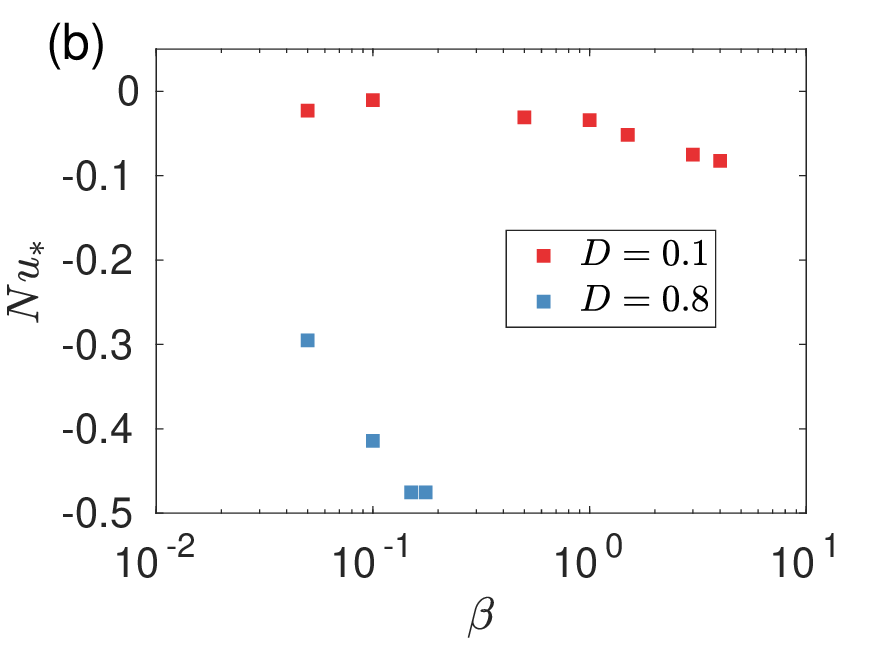}
\caption{\label{fig:renu}  Relative Reynolds (a) and Nusselt numbers (b), $Re_{\ast}$ and $Nu_{\rm ast}$, with respect to the power law exponent $\beta$. The legend indicates the corresponding dissipation number $D$.}
\end{figure}

Finally, we plot the relative Reynolds and Nusselt numbers in Figs. \ref{fig:renu} (a,b).  The relative Reynolds and Nusselt numbers, $Nu_{\ast}$ and $Re_{\ast}$, are given as the ratio of the corresponding difference between the actual value at $\beta\ne 0$ and the one at $\beta= 0$ to that at $\beta= 0$, i.e.,
\begin{equation}
Re_{\ast}=\frac{Re(\beta)-Re(0)}{Re(0)}\,,\quad Nu_{\ast}=\frac{Nu_T(\beta)-Nu_T(0)}{Nu_T(0)}\,.
\end{equation}
From panel (a), we see that for $D= 0.1$, the relative Reynolds number $Re_{\ast}$ increases with $\beta$. We observe a change of up to 40\% for the highest exponent $\beta=4$. This is not surprising since the dynamic viscosity decreases with depth for increasing $\beta$. However, for $D= 0.8$, the relative Reynolds number shows an opposite trend and decreases with $\beta$. This can be related again to the decreased level of density fluctuations with $\beta$ near the top boundary.  

In panel (b) of the same figure, we observe a different trend for the relative Nusselt number $Nu_{\rm ast}$. Despite an increase of $Re_{\ast}$, the relative Nusselt number remains close to zero for growing $\beta$ at $D= 0.1$.  This implies that the turbulent heat transfer efficiency remains the same even under strong variation in thermal conductivity. For the second series at $D= 0.8$, the relative Nusselt number is found to decrease with $\beta$. Under strong stratification conditions, the heat transport efficiency is further decreased due to variable material properties, here by up to 50\%.  In the SSC regime, the major heat transport events are the slender plumes which fall from the top boundary, as already mentioned. Once again, the decreased Nusselt number with $\beta$ seems to correspond to the reduced thermodynamic fluctuations near the top boundary.

\section{Summary and Outlook}
The major motivation of the present study was to extend our previous analysis of compressible turbulent convection with constant material properties in refs. \cite{JPJJSJFM2023,JPJJSPRF2023} to the case with temperature-dependent ones. This leads to non-Boussinesq effects from two combined sources and thus to deviations from the Oberbeck-Boussinesq Rayleigh-B\'{e}nard convection (OB RBC) case: (1) from genuine compressibility and (2) from varying material properties. The latter source was modeled by a power law dependence on the temperature $T$ with a power law exponent $\beta$. We omitted the weaker dependence on pressure $p$. Our present work also extends former numerical studies of a temperature-dependent temperature diffusivity in the Boussinesq approximation by Pandey et al. \cite{pandey2021non,pandey2021nona}. 

Two series of direct numerical simulations were analysed which were obtained at the same Rayleigh and Prandtl numbers, $Ra=10^5$ and $Pr=0.7$, but for different dissipation numbers $D$, a weakly stratified case at $D=0.1$ and a strongly stratified case at $D=0.8$. They were denoted as the OB and SSC regimes in our previous study.\cite{JPJJSPRF2023} The exponents of the corresponding power laws for the thermal conductivity $k(T)$ and the dynamic viscosity $\mu(T)$ could be varied over certain ranges; the larger $\beta$ the stronger the variability of the material parameter in the layer. The same holds for the superadiabaticity $\varepsilon(z)$. The dependence on temperature was chosen such that the mean value of the superadiabaticity remained on average at $\varepsilon_{\rm global}=0.1$ throughout this study.

Our results can be summarized as follows. The mean superadiabatic temperature profiles become increasingly asymmetric with increasing $\beta$. The variation of the specific entropy with height (normalized to the entropy value at the bottom) shows a systematic although weak dependence on $\beta$. The trends close to the top and bottom are however opposite for the SSC regime in comparison to the OB-like regime. It can be traced back to the different couplings of the fluctuations of the thermodynamic state variables, density to temperature for weak and density to pressure for strong stratification, respectively. 

We also detected a $\beta$--dependence for the temperature and density fluctuation profiles, however not a qualitative change of the profiles. The profile of the turbulent Mach number remains nearly unchanged for the weak stratification cases. The degree of compressibility is reduced across the whole layer when $\beta$ is enhanced for the SSC case at $D=0.8$. The turbulent momentum transfer is strongly enhanced for $D=0.1$ and only weakly for $D=0.8$. The turbulent heat transfer remains nearly insensitive for $D=0.1$ and decreases significantly at $D=0.8$ for growing $\beta$. Overall it can thus be stated that $\beta$--dependencies are present for all quantities that we studied; the sensitivity is particularly pronounced for the global momentum and heat transfer. It depends strongly on the degree of stratification which we determine by the dissipation number $D$. We can conclude from our analysis, that the additional inclusion of the temperature dependence of the thermal conductivity and the dynamic viscosity leads to quantitative, but not qualitative changes of the convection dynamics. 

Even though, the present analysis was already very comprehensive, we could show results for one Rayleigh number only. It can be expected that the complexity of parameter dependencies is further increased when a variation of the Rayleigh number is incorporated. Furthermore, we varied both material parameters such that the Prandtl number remains constant. If we would give up this constraint, new convection regime changes can be expected. These two aspects define possible directions for the future research on this subject.

\acknowledgements
It is our great pleasure to dedicate this work to the Special Topic on the occasion of the 75th Birthday of Katepalli R. Sreenivasan. The research of J.P.J. is supported by the Alexander von Humboldt Foundation. The authors gratefully acknowledge the Gauss Centre for Supercomputing e.V. (https://www.gauss-centre.eu) for funding this project by providing computing time through the John von Neumann Institute for Computing (NIC) on the GCS Supercomputer JUWELS at Jülich Supercomputing Centre (JSC). 

\bibliography{all}

\end{document}